\documentclass[twocolumn ,amssymb, amsmath, showpacs, aps, prb]{revtex4} 
\usepackage[dvips]{graphicx}
\begin{document}
\title{Spin-glass-like state in GdCu: role of phase separation and magnetic frustration }
\author{A.Bhattacharyya}
\author{S. Giri}
\author{S. Majumdar}
\email{sspsm2@iacs.res.in} \affiliation{Department of Solid State
Physics,Indian Association for the Cultivation of Science, 2A \& B Raja S. C.
Mullick Road, Kolkata 700 032, INDIA}

\begin{abstract}
We report investigations on the ground state magnetic properties of intermetallic compound GdCu through dc magnetization measurements. GdCu undergoes first order martensitic type structural transition over a  wide temperature window of coexisting phases.  The high temperature cubic  and the low temperature orthorhombic phases have different  magnetic character and they show  antiferromagnetic and helimagnetic orderings below 145 K and 45 K respectively.  We observe clear signature of a glassy magnetic phase below the helimagnetic ordering temperature, which is marked by thermomagnetic irreversibility, aging and memory effects. The glassy magnetic phase in GdCu  is found to be rather intriguing with its origin lies in the interfacial frustration due to distinct magnetic character of the coexisting phases.
\end{abstract}

\maketitle

\section{Introduction}

Spin glass   is an intriguing  example of non-ergodic state marked by several physical properties such as thermomagnetic irreversibility,  slow dynamics, non-exponential decay, magnetic memory effect, etc.~\cite{binder, mydosh, weissman, jonason, lefloch} The non-equilibrium state arises from {\it frustration} due to  competing magnetic interactions among atomic spins  as well as {\it disorder} capable of pining the spins. In recent time, spin glass like non-equilibrium dynamics and time dependent phenomena have been observed in several magnetic systems, where the basic building blocks responsible for the `glassy' behavior are not really the atomic spins, rather spin cluster or bigger spin entity.~\cite{mathieu}  In particular, such behavior have been widely observed in  phase separated  manganites~\cite{maignan, deac, rivadulla, ghivelder, kde} and cobaltites.~\cite{tang, mpatra} The observation of slow dynamics resembling classical spin glass is also extended to  intermetallic alloys~\cite{mkc, sbr, sc} showing first order magnetostructural transition and also to magnetic nanoparticles.~\cite{salamon} Since, the basic building blocks are bigger spin entity, they are often termed as cluster glass, superspin glass or magnetic glass.  
  
\par
The coexisting magnetic phases related to glassyness in manganites and other bulk solids  are often related to first order phase transition (FOPT).~\cite{dagotto} In presence of static disorder, FOPT can lead to coexistence of high-temperature ($T$) parent  and low-$T$ product phases within the region of transition. In certain circumstances, there can also be structural freezing of the parent phase  below the transition point.~\cite{mkc, sbr, pc}  The  glassyness in such phase separated system can have two likely origin.  Firstly, the  slow dynamics of the coexisting phases due to their metastability, and secondly due to frustration arising from the magnetic interaction   between two  clusters having  distinct magnetic nature. The spin glass like state (or often called cluster glass) in phase separated manganites has often been attributed to the coexisting ferromagnetic (FM) and charged order antiferromagnetic (AFM) phases of micrometer size due to FOPT.~\cite{rivadulla, dagotto} It is worth noting that in case of polydispersive non-interacting  nanoparticle system (where the magnetic interaction between two particles is negligible),  a simple model based on the distribution of superparamagnetic relaxation time   can explain the observed glassy magnetic behavior including the memory effect.~\cite{sasaki1, zheng, sasaki2, sdg}

\par
Considering the fact that a large number of   bulk phase separated materials show glassy magnetic behavior, it is pertinent to investigate the role of  inter-cluster magnetic interaction toward the observed state. However, unlike magnetic nanoparticles,~\cite{sasaki2} one can not tune the strength of the magnetic interaction in a bulk material. More importantly, the FOPT in such systems is of magneto-structural type with strong interplay between magnetic and structural degrees of freedom.  As a result,  the structural and the magnetic transitions occur below the same $T$, and their individual role toward the glassyness becomes difficult to differentiate.

\par
Here we report the magnetic investigation on GdCu intermetallic compound, which was reported to show phase coexistence due to first order martensitic transition (MT).~\cite{jcmvd, jab, ho, sathe} The fact that prompted us to investigate GdCu is that  the magnetic and structural transitions  occur at distinctly different $T$.  Therefore, it  provides an opportunity to investigate  the role of magnetic interaction and structural phase separation in determining the ground state magnetic character.

\par
The RCu (R= rare-earth) series of compounds with heavy rare-earth (R $\geq$ Gd) crystallize in the cubic CsCl-type structure (hereafter called C-phase) at high-$T$.~\cite{ai} Some of the members (R = Gd, Tb and Y) show lattice instability and undergo MT at low $T$  to an orthorhombic FeB-type structure (hereafter called O-phase).  GdCu shows long range AFM ordering below $T_N^{C}$ in the C-phase, while it undergoes a second transition below $T_N^{O}$ to a helimagnetic (HM) spin structure in the low-$T$ O-phase. Our investigation shows the existence of  an unconventional  glassy magnetic phase in GdCu below  $T_N^{O}$,  which presumably  arises out of  the magnetic frustration from  coexisting phases.

\section{Experimental Details}
Polycrystalline samples of  GdCu and DyCu were prepared by argon arc melting with the constituent elements of purity better than 99.9  wt\%. The ingots were homogenized at 800$^o$C for 120 h.  Room temperature powder x-ray  diffraction (XRD) patterns were recorded  using a Bruker AXS diffractometer (Cu K$\alpha$ radiation, 2$\theta$ range from 20$^{\circ}$ to 80$^{\circ}$ with a step size of 0.02$^{\circ}$ and 5 s/step counting time). The collected powder patterns were used for phase identification of the given compound using the GSAS software package.~\cite{gsas}  Rietveld refinement data along with XRD pattern of GdCu are  shown in the inset of fig. 1.  Lattice parameter ($a$ = 3.518 \AA) and unit cell volume ($V$ = 43.54 \AA$^3$) are in good agreement with the previously reported data. The analysis indicates that the annealed samples are single phase with cubic CsCl structure  at room temperature.

\par
The $T$ variation of resistivity ($\rho$)  in zero as well as in presence of external magnetic field ($H$)  were measured down to 5 K in a commercial cryogen free high magnetic field system from Cryogenic Ltd., U. K.  Magnetization ($M$) was measured using a commercial Quantum Design SQUID  magnetometer (MPMS XL Ever Cool model).  

\begin{figure}[t]
\vskip 0.4 cm
\centering
\includegraphics[width = 8.5 cm]{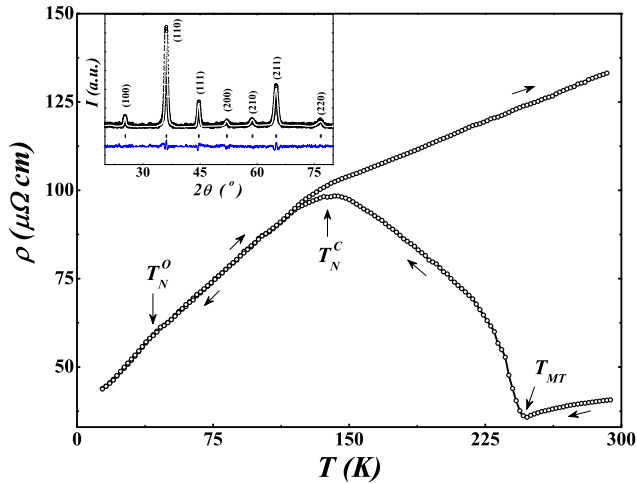}
\caption { (Color online) The main panel shows the zero field resistivity as a function of temperature for cooling and subsequent heating legs for GdCu. The sample was cooled from 300 K down to 5 K and heated back to 300 K again.  The inset shows the Rietveld refinement of  x-ray powder diffraction pattern of GdCu compound. Open circle  represent observed data and the lines  drawn through the data points correspond to the calculated patterns. }
\end{figure}

\section{Results}
Fig. 1  represents the $T$ dependence of $\rho$ of GdCu measured during cooling  and heating. A sharp anomaly is observed around $T_{MT}$ = 250 K in the cooling data  which signifies the MT  from the high-$T$ C-phase to low-$T$ O-phase as reported in previous studies.~\cite{jcmvd, jab, sathe} Only a part of the  thermal hysteresis loop associated with the MT is  visible in the measured $T$ range, because the transition extends up to 650 K in the high-$T$ side.~\cite{jcmvd} In the low-$T$ side, the loop closes at 140 K.~\cite{jab, jcmvd} Similar hysteresis is also present between the field-cooling (FCC) and subsequent  field-cooled heating (FCH) legs of the $M^{-1}$ versus $T$ data (Fig. 2 (a), main panel).  A small anomaly is observed both in the $\rho(T)$ and $M^{-1}(T)$  near  145 K, which matches well with the AFM transition temperature, $T_N^{C}$  of the austenitic C-phase reported earlier.~\cite{jab}  A second magnetic  anomaly  is observed in $\rho(T)$ and $M^{-1}(T)$ data around 45 K (denoted by  $T_N^{O}$) which indicates the onset of helical magnetic ordering of the  martensitic O-phase present in the sample.~\cite{jab} 

\begin{figure}[t]
\vskip 0.4 cm
\centering
\includegraphics[width = 8.5 cm]{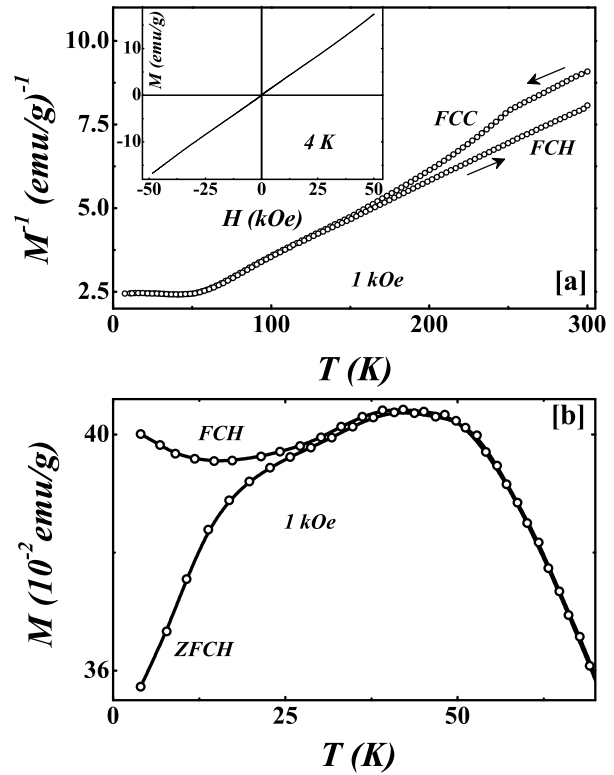}
\caption {(a) Inverse magnetization as a function of temperature measured in the  field-cooling (FCC) and  subsequent field-cooled heating (FCH) protocols for GdCu. The FCC and FCH data show thermal hysteresis down to about 140 K. The inset shows the magnetization as a function of field recorded at  4 K. (b) Zero field-cooled-heating (ZFCH)  and field-cooled-heating (FCH) magnetizations as a function of temperature  in 1 kOe of applied field. }
\end{figure}

\par
In fig. 2 (b), we have shown  $M$ versus $T$ data measured in the zero-field-cooled heating (ZFCH) and FCH protocols with $H$ = 1 kOe.   A clear  bifurcation is observed  between ZFCH and FCH magnetizations [fig. 2 (b)] below $T_N^{O}$, signifying the  existence of thermomagnetic irreversibility associated with the HM transition. The ZFCH magnetization drops with decreasing $T$, while the FCH counterpart shows a sluggish upturn. The feature in the $M(T)$ curve near  $T_N^{O}$ is found to be rather broad. Such broad feature associated with thermomagnetic irreversibility  can be an indication of disordered and/or glassy magnetic phase below  $T_N^{O}$. 

\par
The inset of fig. 2 (a) shows the $M-H$ curve of GdCu measured at 4 K, which  is found to be  linear  without any signature of field induced transition.  The $\rho$ versus $H$ curves  (not shown here) also do not show any signature of metamagnetism. The magnetoresistance (MR = [$\rho(H)-\rho(0)]/\rho(0)$) at 4 K is found to be very small and positive (about 4\% for $H$ = 50 kOe), which is  common  among  bulk AFM  materials.~\cite{yamada} Therefore, one can rule out the possibility of any major role of $H$ in stabilizing different magnetic and structural phases in GdCu. This is possibly indicative of a weak magneto-structural coupling in GdCu unlike manganites~\cite{daf} or magnetic shape memory alloys.~\cite{sc} 

\subsection{Magnetic relaxation}
Considering  magnetic irreversibility in GdCu, the time ($t$) evolution of $M$ was investigated at low $T$. The measurement was performed in two protocols, namely   zero-field-cooled (ZFC) and field-cooled (FC) as described in figs. 3 (a) and (b).  The $t$ dependence of $M$ for both the protocols were measured under 100 Oe of applied field [figs. 3 (a) and (b)].  The ZFC measurement was performed at different temperatures above and below $T_N^{O}$. However, only strong relaxation effect was observed below $T_N^{O}$. The $M(t)$ data collected in the ZFC mode at 30 K ( below $T_N^{O}$) shows 1\% change in 3600 s, while at 60 K (just above $T_N^{O}$) the change is only 0.1\%.  
\par
The time evolution of $M$ have been analyzed on the basis of  various available models of  slow dynamics applicable to magnetic system. It  can often show a power law ($\sim t^{\pm \alpha}$) or a Kohlrausch-Williams-Watt (KWW) stretched exponential  behavior ($\sim \exp{[-(t/\tau)^{ \beta}}]$)~\cite{ito,rvc} where $\tau$ is the characteristic relaxation time and $\beta$ is the shape parameter. We have used both the relations to fit our data, and found that the best fit is obtained with the KWW model. Such model was widely used to analyze the data for spin glass and other disordered magnets.~\cite{phillips} In this  model, $\beta$ lies between 0 and 1 for different  class of disordered materials. The value of $\beta$ was found to be 0.51 and 0.59 for the ZFC and FC data respectively at 4 K. The exponent $\beta$ in the KWW model  signifies the number of intermediate states through which the system should evolve, and it approaches 1 when the number of such intermediate states diminishes.~\cite{xd} Glassy magnetic systems are found to show $\beta$ values over a wide range between 0.2 to 0.6 below the freezing temperature ($T_f$). For example, the spin glass alloy La-Fe-Mn-Si~\cite{wang}  has $\beta$ value close to 0.5 well below $T_f$ , while CuMn (4.0 at \%)~\cite{chu} shows $\beta$ to be between 0.2 to 0.4 depending upon the temperature of measurement.

\begin{figure}[t]
\vskip 0.4 cm
\centering
\includegraphics[width = 8.5 cm]{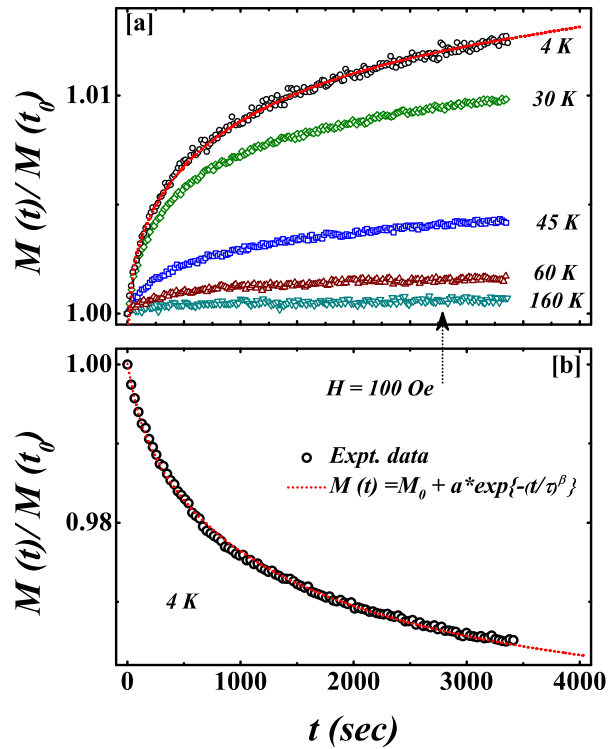}
\caption { (Color online) (a) Time dependence of magnetization at several constant temperatures measured in the zero-field-cooled (ZFC) state with an applied field of 100 Oe for GdCu. The ZFC state is achieved by cooling the sample from room temperature  to desired measurement temperatures in absence of field. (b) Similar measurement in the field-cooled (FC) mode at 4 K, where the FC mode is achieved by cooling the sample from room temperature  to 4 K  in $H$ = 10 kOe  and the time dependence was measured  after reducing the field to 100 Oe. The solid lines through 4 K data (both ZFC and FC modes) are fittings  with stretched exponential function (see text). Here magnetization has been normalized with its initial value at $t_0$ (the staring time for the relaxation measurement).}
\end{figure}

\subsection{Aging}
We  studied the aging effect in GdCu below $T_N^{O}$ (see fig. 4). The sample was first zero -field cooled down to 15 K, and   was allowed to age there for a certain waiting time of  $t_w$.  Subsequently $M$ was measured as a function of time  in presence of   $H$ = 100 Oe. The measurement was performed for three different values of $t_w$, namely 600, 2000 and 5000 s. From the $t$ variation of  $M$ in the resulting  states,  we have calculated  the magnetic viscosity ($S(t) = \frac{1}{H}\frac{\partial M(t)}{\partial \ln {t}}$) as  shown in fig. 4.  It is clear that the $S(t)$ behavior is strongly influenced  by the waiting time.  The $S(t)$ plots show clear peak at the respective $t_w$ values, which is typical experimental evidence of the non-equilibrium states in classical spin glass system.~\cite{mydosh} It is worth noting that such perfect match between $t_w$ and the peak position of $S(t)$ in case of GdCu  is  absent in several glassy magnetic materials indicating a deviation from the classical spin glass behavior (see for example.~\cite{hari, markovich})

\subsection{Memory in  the temperature variation of $M$}
 We probed the  memory in dc magnetization in  GdCu through different modes of measurements.~\cite{jonason, salamon, hari, markovich, sasaki2}   First  we depict the memory in the $T$ dependence of $M$ measured in the FC state following the protocol by Sun {\it et al.}~\cite{salamon} (see fig. 5 (a)).  Chronologically, the measurement was performed in the following steps:  (i) the sample was cooled  in $H$ =  100 Oe  from 100 K to 2 K with intermediate stops  of duration $t_w$ =  1 h each   at $T_{stop}$ =  30, 20 and 10 K. During each stop, $T$ was kept constant, while $H$ was reduced to zero. After each stop, we reapplied $H$ and resumed cooling, which resulted a step-like $M-T$ curve ($M_{FCC}^{stop}$). (ii) After reaching 2 K, the sample was heated back in $H$ = 100 Oe, which produces the so called `memory' curves ($M_{FCH}^{mem}$). (iii) A reference curve  was also recorded by simply allowing the sample to heat in  $H$ = 100 Oe after being field-cooled in the same field without any stop ($M_{FCC}^{ref}$).  The $T$ ramp rate in all the measurement (as well as during field cooling) was kept fixed at 1 K/min. While heating (curve $M_{FCH}^{mem}$  in fig. 5 (a)), we observe striking memory effect signifying nonergodic behavior of the low-$T$ phase.  At each $T_{stop}$, the sample produces clear upturn revealing  the previous history of stops in zero field at that $T$.  This is a  manifestation of memory in the low-$T$ phase.  However, it is interesting to note that {\it no memory was observed above $T_N^{O}$ (i.e. $T_{stop} >T_N^{O}$) }, which once again proves the role of HM ordering for the observed arrested dynamics in GdCu.

\begin{figure}[t]
\vskip 0.4 cm
\centering
\includegraphics[width = 8.5 cm]{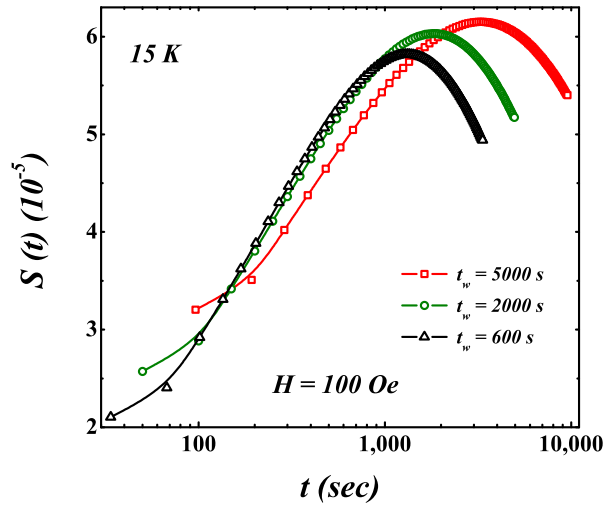}
\caption { (Color online) The waiting time ($t_w$) dependence of magnetic viscosity  for GdCu recorded at 15 K. The relaxation was measured in the zro-field-cooled state at $H$ = 100 Oe  for different values of $t_w$.}
\end{figure}

\par
 We also looked for memory in DyCu, which belongs to the same RCu series of compounds.  However, unlike GdCu, DyCu does not undergo MT and it remains in the C-phase down to low $T$. DyCu undergoes AFM ordering below 63 ~K,~\cite{amara} and shows a second magnetic transition below 20 K. We performed the same memory experiment on DyCu, and no signature of magnetic memory was observed (see fig. 5 (b)). This is in contrast with the behavior of GdCu, and presumably related to the absence of MT related phase coexistence.

\begin{figure}[t]
\vskip 0.4 cm
\centering
\includegraphics[width = 8.5 cm]{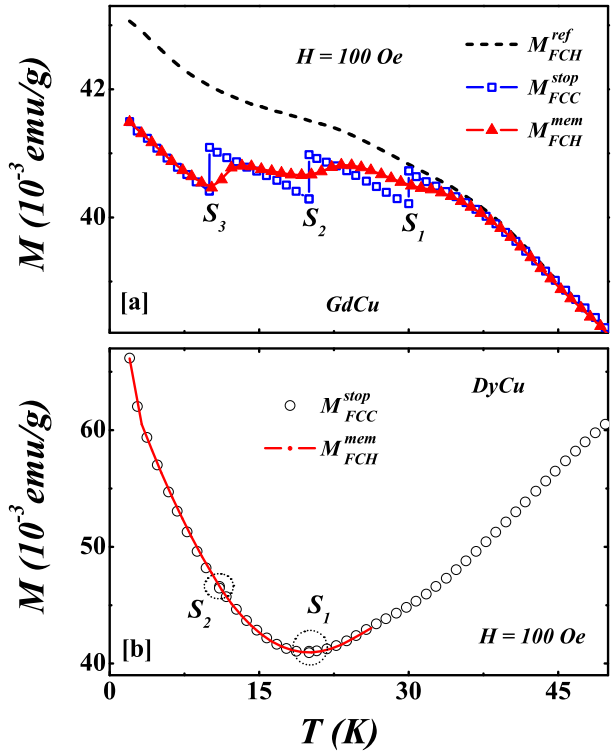}
\caption { (Color online) (a) Field-cooled (FC) memory effect in the temperature ($T$) variation of dc magnetization of GdCu where the curves  were obtained  by cooling the sample in 100 Oe  with intermediate zero-field stops at $T$ = 30 K, 20 K and 10 K ($M_{FCC}^{stop}$) and then subsequent field heating ($M_{FCH}^{mem}$). Reference curve ($M_{FCH}^{ref}$)  was measured on heating  after the sample being field-cooled  in  100 Oe without intermediate stops. (b) shows the similar memory experiment in DyCu.}

\end{figure}

\par
Memory in the FC magnetization can also occur due to a distribution of relaxation time  for a  poly-dispersive  nanoparticle system. GdCu  being a phase separated system, similar effect may also arise from the independent  relaxation of metastable phase clusters. A probable way to rule out such possibility is to investigate the memory in the ZFC magnetization, which is an unequivocal signature of glassy magnetic state originated from cooperative spin-spin interaction.~\cite{sasaki2} Here the protocol is the same as that of FC memory measurement, only the sample was cooled in zero field and  allowed to relax at selected stops (here 20 and 12 K) in zero field only. While  heating in 100 Oe, the sample shows characteristic features at the point of stops. This  is clearer in fig. 6 (b), where we have subtracted the reference ZFC curve ($M_{ZFCH}^{ref}$, measured without stops) from the memory curve ($M_{ZFCH}^{mem}$ measured with stops). The present observation clearly strengthens the view of spin glass like  state in GdCu. Notably, the reference and the `memory' curves coincide exactly at the low and high $T$ end points, which is a signature of the complete eradication of the memory upon heating and cooling.~\cite{markovich}

\subsection{Memory in the time variation of $M$}
The arrested dynamics of the  glassy magnetic phase can be  described by a `free energy landscape picture', which  consists of many local  minima of the free energy separated by a distribution of finite energy barriers.~\cite{chandan} If the system gets trapped in one such local minimum while cooling through the glass transition, the subsequent dynamics is governed by the thermal activation across the energy barriers.  The memory effect and aging are the manifestation of the evolution  from one free energy minimum to another with time.

\par
In order to strengthen the observed memory `dips' in the $M(T)$ data, we  investigated the relaxation behavior with negative $T$ cycling as shown in fig. 7 (a). The relaxation data  was recorded in the ZFC mode. The sample was first cooled down to $T_0$ = 15 K in zero field. Subsequently, $M$ was recorded as a function of $t$ in presence of $H$ = 100 Oe [segment $ab$ in fig. 7 (a)] . After 1 h, the sample was quenched in constant field to a lower $T$, $T_0-\Delta T$ = 10 K and $M(t)$  was recorded further for a time $\sim$ 1 h [segment $cd$]. Finally,  $T$ is turned back to $T_0$ and $M$ was recorded for further 1 h [segment $ef$]. Relaxation process during $ef$ is simply a continuation of the process during $ab$. This memory effect reflects that the state of the system before cooling is recovered when the sample is  cycled back to the initial $T$.  We  also performed simultaneous $H$ and $T$ cycling to check the robustness of magnetic memory in the system [fig. 7 (b)].  Here the protocol is very similar to that of the relaxation experiment depicted in fig. 7 (a). The only difference is that during the temperature quench to  $T_0-\Delta T$ = 10 K, $H$ was also reduced to zero. The continuity of the $M(t)$ curves before and after $T$ and $H$ cycling indicates that the sample is capable of retaining the history even for large change in $M$ (about 10 times).

\begin{figure}[t]
\vskip 0.4 cm
\centering
\includegraphics[width = 8.5 cm]{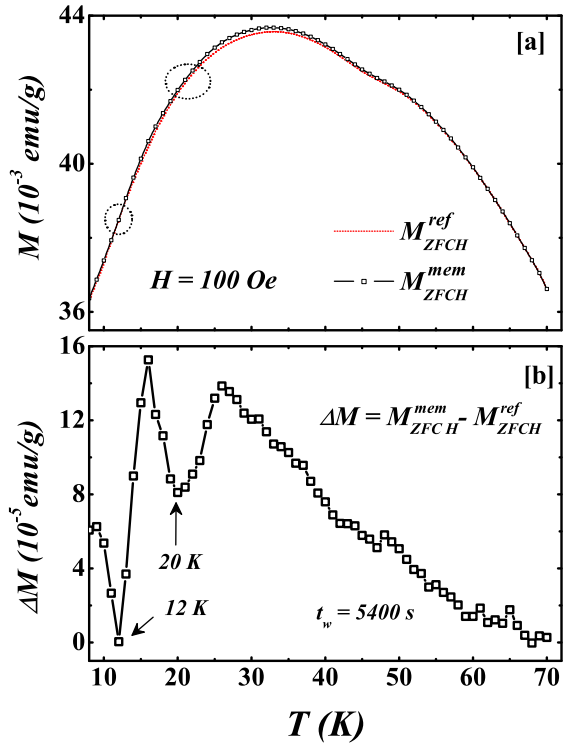}
\caption {(Color online) (a) shows the memory measurement in zero -field cooled condition recorded on  dc magnetization versus temperature data for GdCu. The sample was first cooled in $H$ = 0  down to 4 K, with intermediate stops at 20 K and 12 K for 1.5 h each. The sample was then reheated in $H$ = 100 Oe up to 70 K ($M_{ZFCH}^{mem}$) . A zero field cooled reference curve ($M_{ZFCH}^{ref}$) without intermediate stops during heating is also shown as a dotted line. (b) shows the difference between the magnetization between memory curve (with intermediate stops) and the reference curve.}
\end{figure}

\begin{figure}[t]
\vskip 0.4 cm
\centering
\includegraphics[width = 8.5 cm]{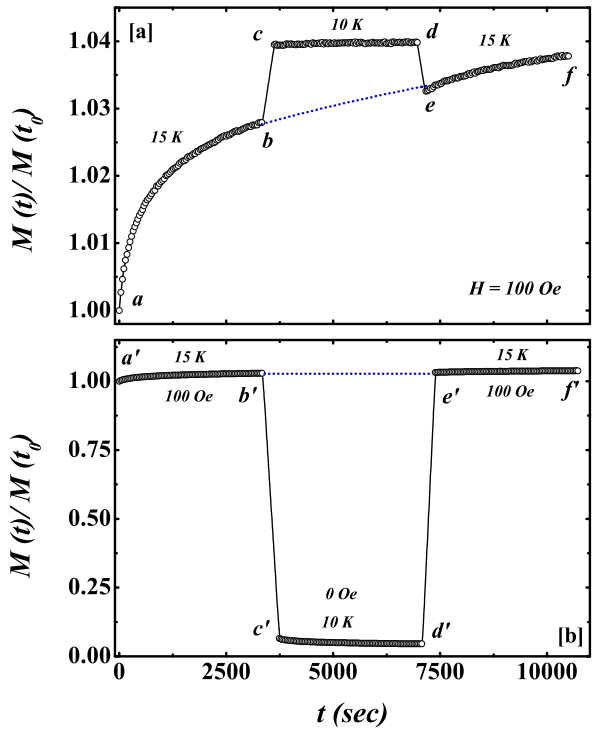}
\caption { (Color online) (a) Magnetic relaxation data at $H$ = 100 Oe in the zero-field-cooled state measured at 15 K ($ab$ and $ef$) along with  an intermediate measurement at 10 K ($cd$) for GdCu. (b) shows a similar measurement, with only exception is that the intermediate 10 K data  was recorded in zero field. Here duration of all the relaxation measurements is $\sim$1 h.}
 \end{figure}

\subsection{Rejuvenation}
It is known that for a nonergodic system, a small positive $T$ cycling can destroy the previous memory and rejuvenates the system.~\cite{jd} We performed such positive $T$ cycling in the relaxation data as shown in fig. 8.  The  difference between the present protocol and that of fig. 7 (a) is that intermittently the sample was heated to a higher $T$, $T_0 +\Delta T$ = 20 K instead of cooling. A sharp contrast is observed between the results of fig. 7 (a) and the present data: the relaxation curves $a''b''$ (before heating) and $e''f''$ (after heating) are found to have different nature and do not continue to posses similar $M$ values. This indicates that heating to a higher $T$ erases the memory and it is  similar to the  rejuvenation observed in various spin glasses and other nonergodic systems.~\cite{pej}

\begin{figure}[t]
\vskip 0.4 cm
\centering
\includegraphics[width = 8.5 cm]{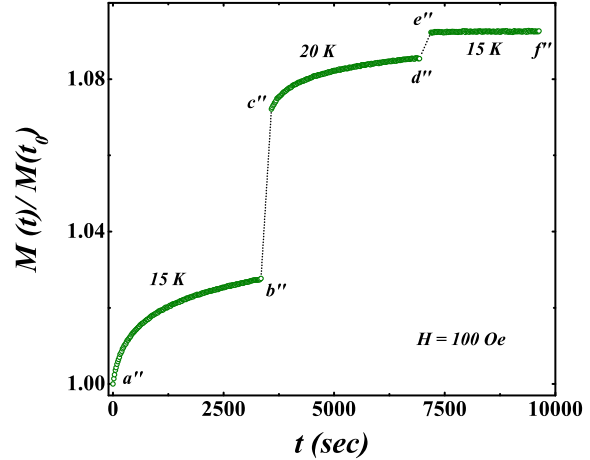}
\caption { (Color online) Magnetic relaxation data measured in the zero field cooled state at 15 K with an intermediate  increases in temperature  from 15 to 20 K for GdCu.  The duration of all the  relaxation studies is $\sim$ 1 h. }
\end{figure}

\section{Discussion}

GdCu is a phase separated system and  in analogy with the well known example of manganites, the glassy phase is likely to be connected to the coexisting phases.  The phase separation in GdCu originates from the FOPT which drives the system from high-$T$ C-phase to low-$T$ O-phase.  The role of FOPT related phase coexistence is evident from our experiment with analogous compound DyCu. The Dy-compound does not undergo FOPT, although like GdCu, it  has  two magnetic transitions. The failure to observe memory in DyCu indicates that low-$T$ glassy magnetic phase is somehow related to the FOPT. In case of GdCu, the thermal hysteresis  is present over a very wide $T$ range, starting from 650 K to 140 K. This is the region of MT, where both O-phase
 and C-phase can coexist. However,  phase coexistence down to very low-$T$ ($\sim$ 4 K) was confirmed by  neutron diffraction experiment. Existence of the high-$T$ C-phase well below the region of MT indicates the {\it  arrested kinetics of the FOPT},~\cite{mkc, sbr, pc, sharma} which allows the metastable C-phase to coexists with the low-$T$ O-phase.

\par
The  C-phase becomes antiferromagnetically ordered below about 145 K, and the slow dynamics  can originate from the \ dependent magnetization of the frozen phase. Arrested dynamics in structurally frozen phase has been observed in Ti-Ni  shape memory alloys.~\cite{shampa} However, our relaxation and memory experiments identify the glassy phase only below $T_N^{O}$ = 45 K, which is the onset point of HM order of the O-phase. Therefore, it is not  the metastability of the structurally frozen phases, rather the cooperative magnetic interaction which decides the glassyness. Otherwise, one would expect the glassy behavior to originate right below $T_ {MT}$ out of the frozen dynamics of the coexisting structural phases.

\begin{figure}[t]
\vskip 0.4 cm
\centering
\includegraphics[width = 7cm]{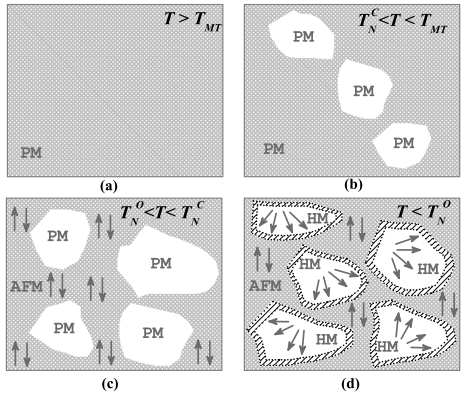}
\caption {A cartoon of the magnetic and structural phase separation in GdCu at different temperatures. The mesh-like region  and the white region denote the C-phase and O-phase respectively.  PM, AFM and HM respectively denote paramagnetic, antiferromagnetic and helimagnetic character of the coexisting phases. The hatched phase boundary in (d) represents interfacial glassy layer due to spin frustration.}
\end{figure}

\par
Our data also  rules out the possibility of a glassy state originating from a distribution of relaxation time  of noninteracting spin clusters, as observed in weakly interacting nanoparticles. Had this been the scenario, we would not expect memory in the ZFC measurement (see fig. 6). Notably, unlike nanoparticles, the magnetic anisotropy is rather weak in a Gd-based system with total orbital quantum number $L$ = 0.

\par
 The  likely mechanism  behind the observed glassyness in GdCu lies in the magnetic correlation. Since it occurs only below  $T_N^{O}$, the low-$T$ HM state plays a deciding role. The glassy phase below  $T_N^{O}$ resembles the classical spin glass state, particularly from the evidence of ZFC memory and aging (figs. 6 and 4 respectively). In GdCu (both in the O and C phases), there is only one crystallographic site~\cite{jab} for Gd which may not be sufficient enough for competing magnetic interaction among atomic spins leading to frustration. In addition, geometrical frustration can be ruled out in a  cubic or orthorhombic crystal structure (both orthogonal). The only possibility that seems viable in the present case is the coexisting structural phases with contrasting magnetic character, which introduces frustration and disorder leading to spin glass like state.
 
 \par
 We propose following scenario for the observed glassyness, which has been depicted in fig. 9. As evident,  GdCu has several magnetic and structural phases. Above $T_{MT}$ (see fig. 1), the system is purely cubic and paramagnetic (PM) [see fig. 9(a)].  According to the  terminology  of MT in metallic alloys, $T_{MT}$ is  called the martensitic start temperature, below which the low-$T$ martensite starts to nucleate. As the system goes below $T_{MT}$, the O-phase develops within the matrix of C-phase, however both are PM in nature. When $T$ is further lowered, the fraction of O-phase increases. Due to arrested kinetics, the complete transformation of the C-phase $\rightarrow$ O-phase does not take place and a fraction of C-phase exists even at the lowest $T$. Now below  $T_N^{C}$, the C-phase orders antiferromagnetically, but the O-phase fraction remains PM, and eventually it orders below  $T_N^{O}$.  Within the temperature region $T_{MT}  <  T_N^{O}$, mixed phase prevails in the sample. However, unless one goes below  $T_N^{O}$, the O-phase remains PM and can not introduce magnetic frustration. Eventually at $T <  T_N^{O}$, both the coexisting phases become magnetically ordered and  interfaces between two phases give rise to frustration owing to their conflicting magnetic character. The effect is analogous  to the surface spin glass state  in interacting $\gamma$-Fe$_2$O$_3$ nanoparticles~\cite{martinez} with phase boundaries in GdCu playing the part of surface layers of the nanoparticles. The interesting point seen in case of GdCu is that  the spin glass like state does not originates right below the first order transition, rather it develops below the low-$T$ HM transition temperature point. The presence of C-phase and O-phase clusters may give rise to a frozen local strain order as observed in Ti-Ni alloys~\cite{shampa}, but that does not gives rise to a  magnetic glassy phase as long as $T >T_N^{O}$. It is to be noted that previous heat capacity study~\cite{ho} showed broad peak around HM transition point, which is also a signature of the onset of glassy phase in a material.

 \par
In conclusion, we observe glassy magnetic phase in GdCu marked by slow dynamics and clear memory effect.  Our results suggest that cooperative magnetic correlation between the coexisting phases is the primary cause of the observed spin glass like state. Such understanding can be quite useful in interpreting the spin glass like state in other phase separated systems including manganites.

\section{Acknowledgment} 
AB would like to thank Council for Scientific and Industrial Research (India) for financial support.

\end{document}